\documentclass[sigconf]{acmart}
\AtBeginDocument{%
  \providecommand\BibTeX{{%
    \normalfont B\kern-0.5em{\scshape i\kern-0.25em b}\kern-0.8em\TeX}}}

\usepackage{booktabs,makecell, multirow, tabularx, enumitem}

\newcommand{\ourmethod}[1]{}
\renewcommand{\ourmethod}[1]{\texttt{HOFA}}

\usepackage{adjustbox}

\setcopyright{acmcopyright}
\copyrightyear{2018}
\acmYear{2018}
\acmDOI{XXXXXXX.XXXXXXX}

\acmConference[Conference acronym 'XX]{Make sure to enter the correct
  conference title from your rights confirmation emai}{June 03--05,
  2018}{Woodstock, NY}
\acmBooktitle{Woodstock '18: ACM Symposium on Neural Gaze Detection,
 June 03--05, 2018, Woodstock, NY} 
\acmPrice{15.00}
\acmISBN{978-1-4503-XXXX-X/18/06}

\begin{document}

\settopmatter{printacmref=false} 
\renewcommand\footnotetextcopyrightpermission[1]{} 

\title{\ourmethod{}: Twitter Bot Detection with Homophily-Oriented Augmentation and Frequency Adaptive Attention}



\author{Sen Ye}
\email{ys2003@stu.xjtu.edu.cn}
\affiliation{%
  \institution{Xi'an Jiaotong University}
  \city{Xi'an, Shaanxi}
  \country{China}}

\author{Zhaoxuan Tan}
\email{tanzhaoxuan@stu.xjtu.edu.cn}
\affiliation{%
  \institution{Xi'an Jiaotong University}
  \city{Xi'an, Shaanxi}
  \country{China}}

\author{Zhenyu Lei}
\email{fischer@stu.xjtu.edu.cn}
\affiliation{%
  \institution{Xi'an Jiaotong University}
  \city{Xi'an, Shaanxi}
  \country{China}}

\author{Ruijie He}
\email{MCPlayer542@stu.xjtu.edu.cn}
\affiliation{%
  \institution{Xi'an Jiaotong University}
  \city{Xi'an, Shaanxi}
  \country{China}}

\author{Hongrui Wang}
\email{wanghongrui@stu.xjtu.edu.cn}
\affiliation{%
  \institution{Xi'an Jiaotong University}
  \city{Xi'an, Shaanxi}
  \country{China}}

\author{Qinghua Zheng}
\email{qhzheng@mail.xjtu.edu.cn}
\affiliation{%
  \institution{Xi'an Jiaotong University}
  \city{Xi'an, Shaanxi}
  \country{China}}

\author{Minnan Luo}
\email{minnluo@xjtu.edu.cn}
\affiliation{%
  \institution{Xi'an Jiaotong University}
  \city{Xi'an, Shaanxi}
  \country{China}}







\renewcommand{\shortauthors}{Trovato and Tobin, et al.}

\begin{abstract}
Twitter bot detection has become an increasingly important and challenging task to combat online misinformation, facilitate social content moderation, and safeguard the integrity of social platforms. Though existing graph-based Twitter bot detection methods achieved state-of-the-art performance, they are all based on the homophily assumption, which assumes users with the same label are more likely to be connected, making it easy for Twitter bots to disguise themselves by following a large number of genuine users.
To address this issue, we proposed \ourmethod{}, a novel graph-based Twitter bot detection framework that combats the heterophilous disguise challenge with a homophily-oriented graph augmentation module (\textsc{Homo-Aug}) and a frequency adaptive attention module (\textsc{FaAt}). Specifically, the \textsc{Homo-Aug} extracts user representations and computes a $k$-NN graph using an MLP and improves Twitter's homophily by injecting the $k$-NN graph. For the \textsc{FaAt}, we propose an attention mechanism that adaptively serves as a low-pass filter along a homophilic edge and a high-pass filter along a heterophilic edge, preventing user features from being over-smoothed by their neighborhood. We also introduce a weight guidance loss to guide the frequency adaptive attention module. Our experiments demonstrate that \ourmethod{} achieves state-of-the-art performance on three widely-acknowledged Twitter bot detection benchmarks, which significantly outperforms vanilla graph-based bot detection techniques and strong heterophilic baselines. Furthermore, extensive studies confirm the effectiveness of our \textsc{Homo-Aug} and \textsc{FaAt} module, and \ourmethod{}'s ability to demystify the heterophilous disguise challenge.

\end{abstract}

\maketitle

\section{Introduction}
Twitter, a widely-used online social media platform, has become an integral part of people's daily lives for sharing information and communicating with one another. Unfortunately, the integrity of Twitter is being compromised by accounts controlled by automated programs, which are known as \textit{Twitter bots}. Twitter bots are posing threats to online social platforms by spreading misinformation \cite{misinformation1, misinformation2, misinformation3, cresci2023demystifying}, interfering in the elections in the United States and Europe \cite{ferrara2017disinformation, detectingpoliticalbots}, propagating extreme ideology \cite{berger2015isis, sarmiento2022identifying}, and promoting conspiracy theories \cite{2020covid, onlineConspiracy}. Consequently, there is an urgent need for effective Twitter bot detection methods to identify Twitter bots and mitigate their negative influences.

To win the arms race with bot manipulators, Twitter bot detection researchers proposed feature-, text-, and graph-based detection methods, where users' metadata, text information, and follower/following relationship are leveraged to identify bot accounts. Feature-based methods extract features from user metadata \cite{sgbot, lee2011seven}, user timeline \cite{mazza2019rtbust}, and follow relationship \cite{botrgcn, rgt} and feed them into a traditional classifier such as Random Forest for bot detection. 

\begin{figure}[t]
    \centering
    \includegraphics[width=0.95\linewidth]{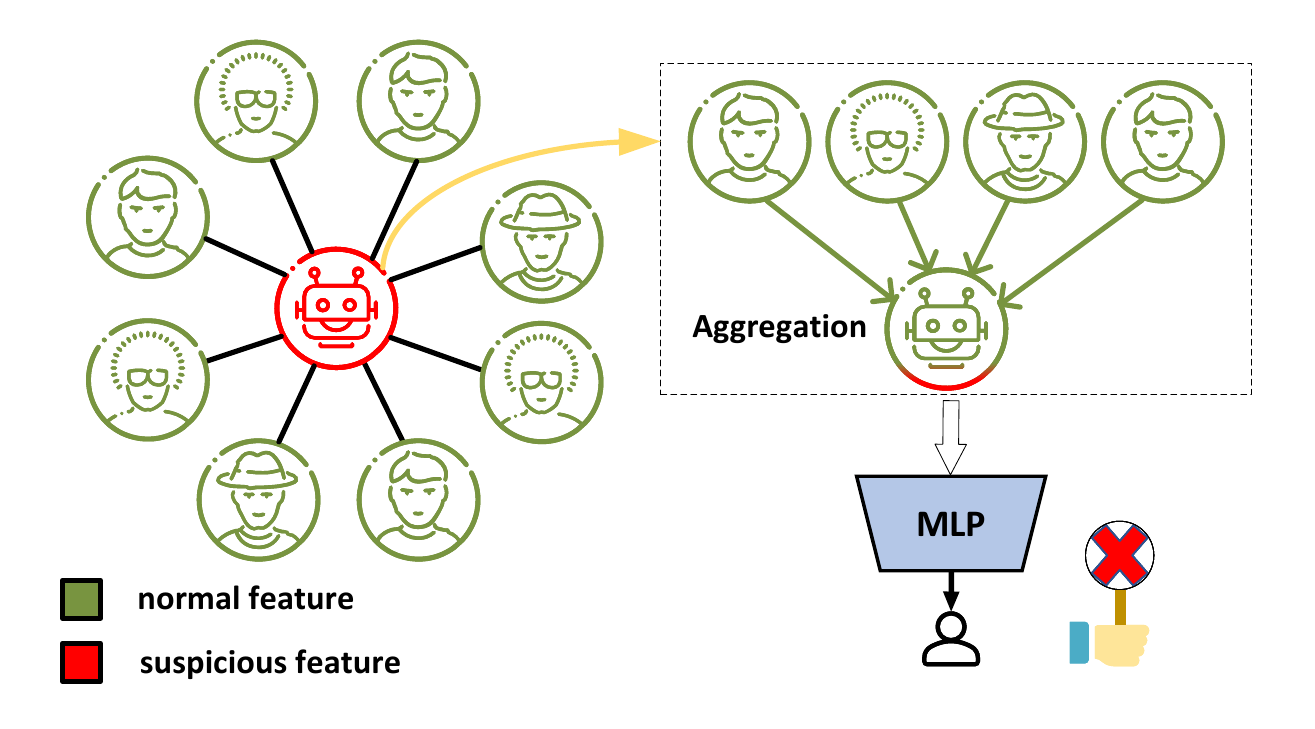}
    \caption{An example of heterophilous disguise where a bot account follows a vast majority of genuine users and the suspicious features are smoothed by neighboring genuine users' normal features.}
    \label{fig:teaser}
\end{figure}

However, bot manipulators often deliberately tamper with hand-crafted features to evade feature-based detection methods \cite{cresci2017paradigm,cresci2020decade}.
Researchers also proposed text-based methods, where NLP techniques such as word embeddings \cite{quezada2019lightweight,wordembedding}, Recurrent Neural Networks \cite{metadatafeature, diaz2020integrated, diaz2022language}, and pretrained language models \cite{bert, usingbert} are leveraged to encode tweet content and identify malicious intent. However, text-based methods fall short when encountering novel Twitter bots since they intersperse malicious tweets with normal tweets stolen from genuine users \cite{cresci2020decade}. With the recent advances of graph neural networks, researcher developed graph-based Twitter bot detection methods, which interpret Twitter as a graph and user nodes are connected by follower/following relationships. Then Graph Neural Network (GNN) such as GCN \cite{gcn, ali2019detect}, RGCN \cite{rgcn, botrgcn}, and RGT \cite{rgt} is employed to learn user representation for bot detection. Graph-based methods achieve state-of-the-art performance and have been shown to be more effective in detecting novel bots \cite{botrgcn, rgt} and generalization \cite{twibot22} compared with feature- and text-based methods.


Although significant progress has been made in graph-based Twitter bot detection methods, they are vulnerable to \textit{heterophilous disguised} bot accounts. Existing graph-based bot detection methods are built upon the homophily assumption, \emph{i.e.}, users (nodes) with similar features or the same label are prone to be connected, and use a low-pass filter to smooth user features within the neighborhood. This assumption smoothes the representations of connected users, making it easy for bot accounts to evade detection by following genuine users.
Figure \ref{fig:teaser} illustrates the heterophilous disguise, by following a large number of genuine users, the suspicious features of a bot account can be smoothed by its human neighborhood and therefore evade the homophilic graph-based detection. 

\begin{figure}[t]
    \centering
    \includegraphics[width=0.95\linewidth]{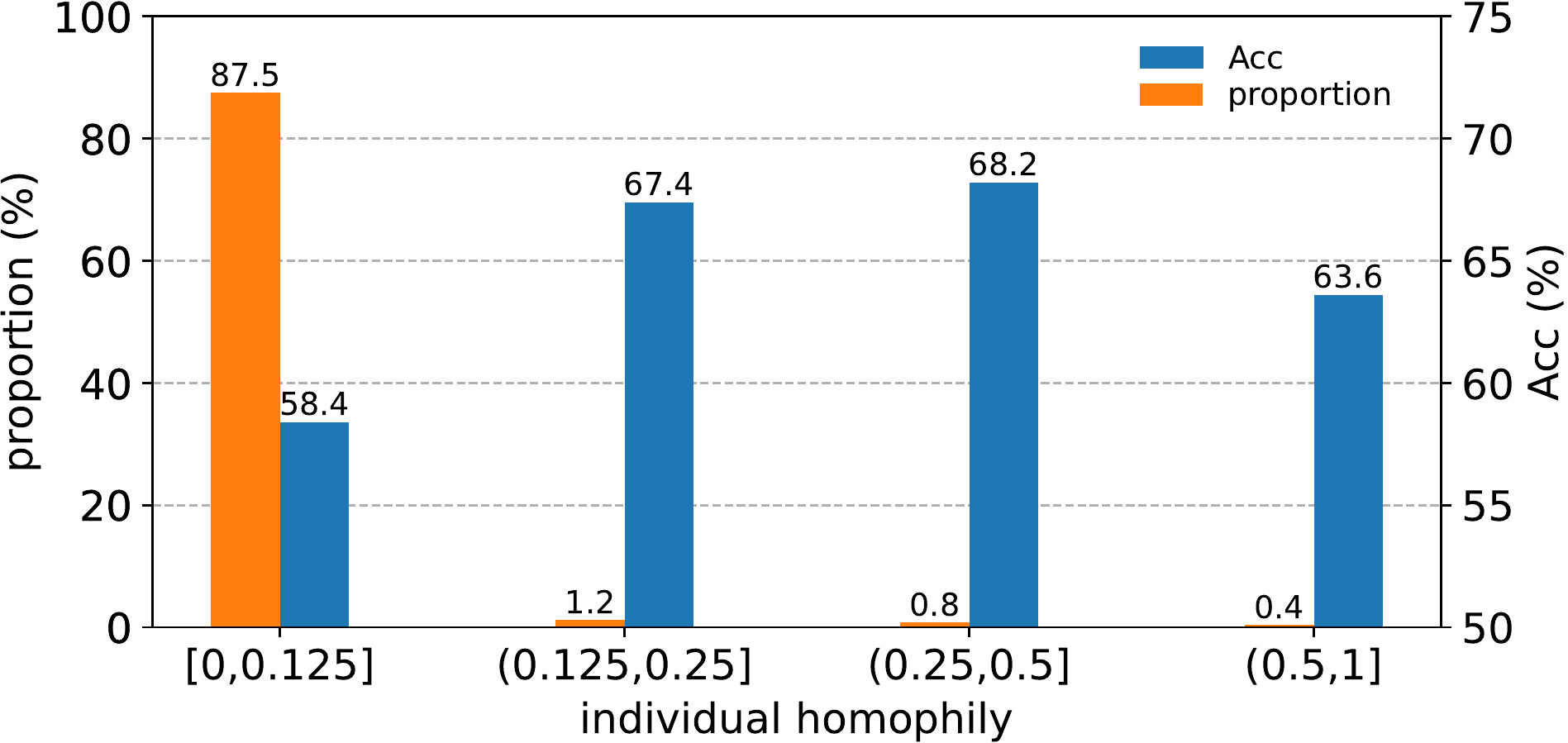}
    \caption{Distribution of bots' individual homophily and GCN's accuracy on MGTAB-22 \cite{mgtab}.}
    \label{fig:dataset_h}
\end{figure}

To validate the heterophilous disguise challenge, we calculate the individual homophily score (the ratio of users with the same label in their neighborhood) for each bot in MGTAB-22 \cite{mgtab} dataset and categorize these bots into 4 groups according to their homophily score (bots without neighbors in the dataset are excluded). We further test GCN's bot detection accuracy for these 4 groups of users and present the results in Figure \ref{fig:dataset_h}. The orange bars show that a large portion of bot accounts' homophily scores are low, indicating the common existence of heterophilous disguise challenge. The blue bars show that bot detection performance is strongly relevant to homophily scores, where the detection performance of users with low homophily scores is lower than that with high homophily scores.

To address the challenges of heterophilous disguise in Twitter bot detection, we propose \ourmethod{}, a novel Twitter bot detection framework that combats heterophilous disguise with a homophily-oriented augmentation module (\textsc{Homo-Aug}) and a frequency adaptive attention module (\textsc{FaAt}). For homophily-oriented augmentation module, we first train an MLP to classify nodes in the training set. Armed with this MLP, we extract hidden representations for each node and then compute the $k$-nearest neighbor set for each node based on the cosine similarity to form the $k$-NN graph \cite{eppstein1997nearest}. We then augment the original graph by injecting $k$-NN graph therefore introducing more homophilic information in the graph. For the frequency adaptive module, attention weights are computed to capture both low- and high-frequency information in heterophilous settings. Different from vanilla attention where a $\mathrm{softmax}$ is applied to map attention weights to $[0,1]$, we use $\textit{tanh}$ to map the attention weights to range $[-1,1]$, where positive attention weights correspond to low-pass filters and negative weights correspond to high-pass filters \cite{fagcn}. To adaptively filter node features, we use a low-pass filter along homophilic edges to smooth features of nodes of the same label. In contrast, a high-pass filter is desired to apply along a heterophilic edge to sharpen the representations of nodes of different labels. To boost the performance and \ourmethod{}'s robustness, we further introduce an auxiliary loss to guide the edge attention weight learning. 

Experimental results demonstrate that \ourmethod{} achieves state-of-the-art performance on three Twitter bot detection benchmarks by significantly outperforming existing homophilic graph-based Twitter bot detection methods and strong heterophilic graph learning baselines. Extensive studies also bear out the effectiveness of our proposed homophily-oriented augmentation, frequency adaptive attention module, and \ourmethod{}'s superiority in solving the heterophilic disguise challenge. To summarize, our contributions are as follows:
\begin{itemize}[leftmargin=*]
    \item We have identified a new challenge in Twitter bot detection: heterophilous disguise. After investigating existing graph-based Twitter bot detection benchmarks, we have ascertained the prevalence of this challenge and its existence imposes a significant hindrance to the effectiveness of graph-based methods.
    \item As the first effort to address the heterophilous disguise challenge in bot detection, we propose \ourmethod{} consisting of a homophily-oriented graph augmentation module (\textsc{Homo-Aug}) improving the homophily of the original graph and a frequency adaptive attention module (\textsc{FaAt}) to adaptively combine low- and high-frequency information to learn better node representations in heterophilous graph. 
    \item Armed with \ourmethod{}, we achieve state-of-the-art performance on three widely acknowledged Twitter Bot Detection benchmarks. Results demonstrate that \ourmethod{} outperforms all Twitter bot detection baselines as well as strong heterophilic graph learning baselines. Further experiments illustrate the effectiveness of the proposed homophily-oriented graph augmentation module and the frequency adaptive attention module. 
\end{itemize}

\section{Related Work}
\subsection{Twitter Bot Detection}
Existing Twitter bot detection methods mainly fall into three categories: feature-, text-, and graph-based. 

\textit{Feature-based} methods conduct feature engineering by designing discriminative features with user metadata \cite{metadatafeature}, tweets \cite{tweetfeature}, temporal patterns \cite{temporalfeature}, descriptions \cite{descriptionfeature}, and follow relationships \cite{satar}. Then features are fed into a traditional classifier such as Support Vector Machine (SVM) \cite{traditionsvm}, Naive Bayes \cite{traditionbayes}, and Random Forest \cite{traditionrf} to identify bot account. Several unsupervised methods are proposed to discover the underlying patterns of social bots, such as clustering \cite{traditioncluster} and anomaly detection \cite{botwalk}. However, bot manipulators are increasingly aware of these feature-based detectors and can deliberately disguise themselves by fabricating bot features \cite{cresci2020decade, satar}.

\textit{Text-based} methods use techniques from natural language processing to process user descriptions and tweets for bot detection. These methods encode users' tweets and descriptions with word embeddings \cite{wordembedding}, recurrent neural networks \cite{metadatafeature}, attention mechanism \cite{satar}, and pretrained language models \cite{bert} to conduct bot detection. Further researches focus on incorporating user features \cite{tweet_metadata}, learning unsupervised user representations \cite{satar}, and addressing multi-lingual issues \cite{knauth2019language}.
However, novel bots are increasingly stealing text content from genuine users \cite{cresci2020decade}, posing a challenge to text-based methods. Furthermore, \citet{botrgcn} have found that relying solely on text content is not robust or accurate enough.

\textit{Graph-based} methods envision Twitter as a graph where users are connected by follower/following relationship and leverage graph machine learning methods for bot detection. \citet{dehghan} exploit typical graph features such as node centrality, closeness centrality, and \textit{etc}. \citet{graphhist} use graph latent local structure and detect bot accounts based on a multi-channel histogram. \citet{ali2019detect} uses graph neural networks (GNNs) to detect bots. Recently, many methods consider the intrinsic heterogeneity in the Twitter network and employ heterogeneous graph neural networks \cite{botrgcn, rgt}, achieving state-of-the-art performance. In addition, many works exploit multi-modality user information to fight against disguised bots. \citet{bic} propose BIC to facilitate text-graph interaction. \citet{botmoe} propose BotMoE which constructs modal-specific experts to jointly utilize information from metadata, text, and graph modality. \citet{tan2023botpercent} construct the largest ensemble to probe the bot percentage on Twitter.


Though much research progress has been made, existing graph-based Twitter bot detection methods are based on the homophily assumption, that is, assuming that bot accounts follow other bots and humans follow other humans. However, we are the first to find this assumption to be inapplicable in modern Twitter bot detection scenarios, and thus propose \ourmethod{} to address the heterophilous challenge in this work.

\subsection{Heterophilous Graph Learning}

Graph Neural Networks (GNNs) have demonstrated exceptional performance in various graph-based machine learning tasks \cite{gcn, sage, gat}. Despite the abundant designs of GNN architectures, most of them implicitly follow the homophily assumption \cite{h2gcn, linkx}. Homophilic GNNs act as low-pass filters \cite{gcn, fagcn} and smooth features along the graph topology, resulting in similar predictions for neighboring nodes. However, many real-world graphs exhibit heterophily that linked nodes have dissimilar features and different class labels, which is the opposite of homophily \cite{h2gcn, linkx, pandit2007netprobe}. Many researches have found that homophilic GNNs fail to learn discriminative features in non-homophilous settings \cite{mixhop, fagcn}. 

To address the heterophily challenge, researchers have proposed various approaches. \citet{mixhop} identify the challenge of learning on heterophilous graphs using homophilic GNNs and propose MixHop to extract features from multi-hop neighbors to obtain more homophilic information. Geom-GCN \cite{geom-gcn} handles heterophily by precomputing unsupervised node embeddings and defining a bi-level aggregation process. LINKX \cite{linkx} is a simple and scalable method that works by separately embedding the adjacent matrix and node feature and combining them using simple MLP.
H2GCN \cite{h2gcn} addresses heterophily through three key designs, namely ego and neighbor feature separation, higher-order neighbors aggregation, and combination of intermediate representations. \citet{fagcn} propose frequency adaptive graph filters that learn edge-level aggregation weights, which can be negative. CPGNN \cite{cpgnn} models label correlations through a compatibility matrix and propagates a prior belief estimation through this matrix, which can learn discriminative representations on heterophilic graphs.

While a considerable amount of researches has been conducted in the domain of heterophilous graph learning, its application in the context of social networks has yet to be explored. Drawing inspiration from the progress made in heterophilous graph learning, we present \ourmethod{}, which aims to tackle the challenge of heterophilous disguise in Twitter bot detection.

\section{Preliminaries}
Let $\mathcal G=(\mathcal V, \mathcal E, A, \mathcal R)$ denotes a Heterogeneous Information Network (HIN) derived from Twitter, where $\mathcal{V}$ is the set of users with $N = \lvert \mathcal V\rvert$; $\mathcal E$ is the edge set without self-loops; $A \in \mathbb{R}^{N \times N}$ is the symmetric adjacency matrix with $A_{i,j} = 1$ if $e_{i,j} \in \mathcal E$, otherwise $A_{i,j} = 0$; $\mathcal R$ is the set of edge types. The neighbor set of the user $v \in \mathcal V$ is $\mathcal N(v)$ and the neighbor set of $v$ specified to edge type $r \in \mathcal R$ is denoted by $\mathcal N_r(v)$. The users' features are represented by $\mathbf{X} \in \mathbb{R}^{N \times F}$, where $F$ is the number of features for each user. 


\textit{Measurement of Homophily \& Heterophily.}
The homophily characterizes how likely the nodes with the same label are connected in a graph. Generally, there are three widely used metrics for measuring homophily: node homophily \cite{geom-gcn}, edge homophily \cite{h2gcn}, and class insensitive edge homophily \cite{linkx}. 
The node homophily is the average fraction of the neighbors with the same label of each node:
\begin{align}
    h_{\text{node}} = \frac{1}{N} \sum_{v \in \mathcal V} \frac{\lvert \{ u \in \mathcal N(v): y_u=y_v \} \rvert}{\lvert \mathcal N(v) \rvert},
\end{align}
The edge homophily is the fraction of edges which connect nodes with the same label:
\begin{align}
    h_{\text{edge}} = \frac{\lvert \{ (u,v) \in \mathcal E: y_u=y_v \} \rvert}{\lvert \mathcal E \rvert},
\end{align}
The edge homophily is further improved to be insensitive of the number of classes and the size of each class:
\begin{align}
    h = \frac{1}{C-1} \sum_{k=1}^{C} \max(0, h_k - \frac{\lvert C_k \rvert}{\lvert 
    \mathcal V \rvert})
\end{align}
where $C$ is the number of classes, $\lvert C_k \rvert$ is the number of nodes of class $k$ and $h_k$ is the edge homophily of nodes of class $k$. We use the class insensitive homophily $h$ as measurement to reduce the effect of class imbalance issue in this paper. The range of $h$ is $[0,1]$. Graphs with strong homophily have higher values of $h$ closer to $1$ and graphs with strong heterophily have lower values of $h$ closer to $0$. Specifically, for graphs that edges are independent of node labels, $h = 0$ \cite{linkx}.

\begin{figure*}
\centering
\includegraphics[width=0.9\linewidth]{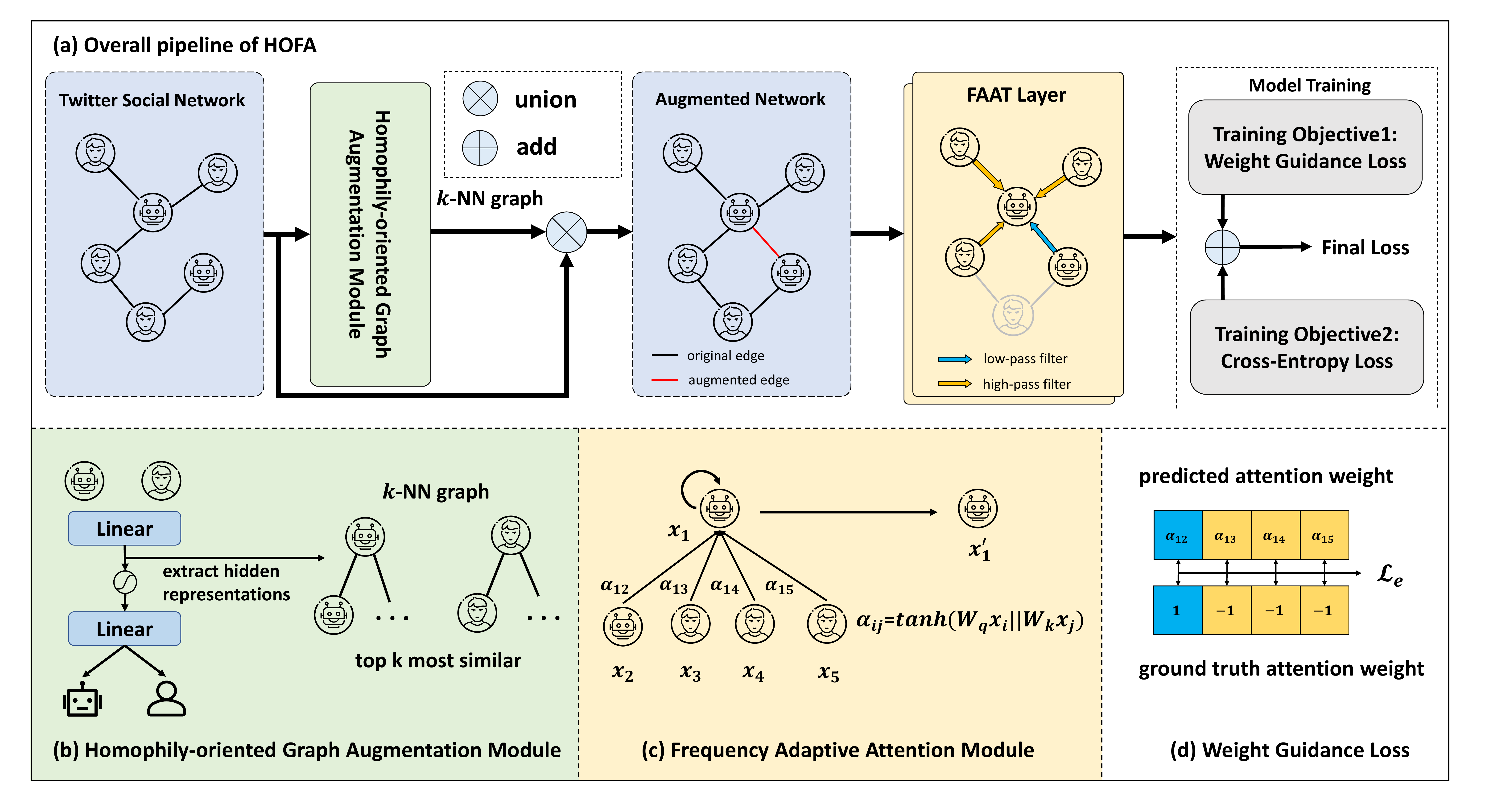}
\caption{\ourmethod{}, a unified framework to combat the heterophilous disguise challenge in Twitter bot detection. The graph structure of Twittersphere is first augmented by homophily-oriented graph augmentation (\textsc{Homo-Aug}), then the frequency adaptive attention (\textsc{FaAt}) is employed to adaptively combine low- and high-frequency information. The model is optimized with the summation of weight guidance loss and cross entropy loss.}
\label{fig:overview}
\end{figure*}

\section{Methodology}
In this section, we will first introduce a novel and simple homophily-oriented graph augmentation module (\textsc{Homo-Aug}) to improve the homophily of heterophilous graphs. Then we will illustrate the frequency adaptive attention module (\textsc{FaAt}) to adaptively capture the low- and high-frequency information in the Twitter network to learn discriminative representations for heterophilous disguised accounts. The overview of \ourmethod{} is shown in Figure \ref{fig:overview}.

\subsection{Homophily-Oriented Graph Augmentation}
Graph data augmentation aims at improving the performance and enhancing the robustness of deep graph models. Previous graph data augmentation methods include edge perturbation, node dropping, attribute masking, and graph sampling \cite{data_aug_survey, GCL}. Among these methods, edge perturbation has proven to be more effective for social networks \cite{GCL}. 
However, randomly adding or removing edges fails to contribute to heterophilous graph learning, as it does not enhance the graph's homophily. Consequently, the subpar performance of vanilla GNNs persists.

To boost the heterophilous graph learning, we propose a simple method \textsc{Homo-Aug} to improve the homophily of graph data by injecting homophilic edges. The proposed method is two-fold, we first train a two-layer MLP for classification and extract the MLP-based representations for all users, we then calculate the $k$-nearest neighbor graph \cite{eppstein1997nearest} using cosine similarity. By injecting the $k$-NN graph into the original graph, \textsc{Homo-Aug} can effectively improve the graph homophily in the Twitter network. For more specific steps, we first train a two-layer MLP with labels in the training set utilizing cross entropy loss to obtain the user representations:
\begin{align}
    \hat{\mathbf{y}} = \mathrm{softmax}(\sigma (\mathbf{XW_1})\mathbf{W_2}),
\end{align}
where $\mathbf{W_1}$ and $\mathbf{W_2}$ are learnable weight matrixs and $\sigma$ is the non-linear activation function (e.g. LeakyReLU \cite{leakyrelu}). Since the MLP does not utilize the original graph structure, MLP-based user representation remains unaffected by the heterophily of the original graph. Next, we compute a $k$-nearest neighbor graph based on the similarity of the hidden representations extracted by the MLP. The hidden representation for each user $u$ is given by:
\begin{align}
    \mathbf{h}_u = \mathbf{W_1x}_u.
\end{align}
After extracting user representations with MLP, we then calculate the $k$-NN graph, where the nearest neighbor set for user $u$ is:
\begin{align}
    \mathrm{k}\text{-NN}(u) = \mathrm{TopK}_v \cos (\mathbf{h}_u, \mathbf{h}_v) \qquad \text{for } v \in \mathcal{V}\setminus\{u\},
    \label{eq:knn}
\end{align}
where $\cos(\cdot, \cdot)$ denotes cosine similarity.
We augment the Twitter network by amalgamating the original edge set with the $k$-NN edge set, resulting in an augmented graph $\tilde{E}$. To maintain the integrity of the original graph structure, we designate a new edge type for the added edges.

\subsection{Frequency Adaptive Attention}

Notably, despite \textsc{Homo-Aug} can augment graph homophily, the heterophilous disguise challenge in the Twitter network may persist and superfluous edges between users may be introduced. To mitigate these negative impacts as well as the intrinsic heterophily in Twittersphere, we introduce \textsc{FaAt}. 

Previous graph-based Twitter bot detection methods are based on homophily assumption and serve as low-pass filters, that being said, making the representations of users along an edge become similar \cite{fagcn}, which works well on homophilic graphs. However, our study reveals the intrinsic heterophily in the Twitter network (heterophilous disguise challenge), causing the traditional homophilic graph-based methods to fall short. For instance, the representations of users with different labels would be confused when connected with heterophilic edges. To address the heterophilous disguise challenge, we propose \textsc{FaAt}, which adaptively chooses low-pass and high-pass filters along user connections using edge weights ranging from -1 to 1. Negative edge weights indicate high-pass filters applied along heterophilic edges to sharpen neighboring user representations and make them more discriminative, while positive edge weights correspond to low-pass filters applied along homophilic edges to smooth neighboring user representations. Additionally, we propose an auxiliary loss called \textsc{WeGL} to guide edge weights learning. Here comes the specific steps, we first concatenate users' description, numerical, and categorical features in:


\begin{align}
    \mathbf{x}_{u}^{(0)} = \sigma (\mathbf{W}^{(0)}[\mathbf{x}_d\Vert\mathbf{x}_\textit{num} \Vert \mathbf{x}_\textit{cat}]),
\end{align}
where $\mathbf{W}^{(0)}$ is the learnable weight matrix, $\sigma$ denotes the non-linear activation function, $[\cdot|\cdot]$ means concatenation operation, $\mathbf{x}_d$, $\mathbf{x}_\textit{num}$, and $\mathbf{x}_\textit{cat}$ respectively denote users' description embeddings, numerical features, and categorical features adopted from \citet{botrgcn}.

\textbf{Frequency Adaptive Attention Coefficient.}
For each edge $e_{uv} \in \mathcal{E}$, \textsc{FaAt} learns a coefficient $\alpha_{uv}$ within range $[-1,1]$ to to function as an adaptive low- or high-pass filter. Negative values indicate low-pass filters while positive values indicate high-pass filters \cite{fagcn}. This is based on the central user feature $\mathbf{x}_u$ and the neighbor user feature $\mathbf{x}_v$. In the $l$-th layer, we first obtain the query and key vector from $\mathbf{x}_u^{(l-1)}$ and $\mathbf{x}_v^{(l-1)}$:
\begin{gather}
    \mathbf{q}_{u}^{(l)} = \sigma (\mathbf{W}_q^{(l)} \mathbf{x}_u^{(l-1)}), \\
    \mathbf{k}_{v}^{(l)} = \sigma (\mathbf{W}_k^{(l)} \mathbf{x}_v^{(l-1)}),    
\end{gather}
where $\mathbf{W}_q^{(l)}$ and $\mathbf{W}_k^{(l)}$ are learnable weight matrix for the $l$-th layer, we then compute the frequency adaptive attention coefficient $\hat\alpha_{uv}^{(l)}$ as follows:
\begin{align}
     \hat\alpha_{uv}^{(l)} = \tanh(\mathbf{g}^{T}[\mathbf{q}_{u}^{(l)}\Vert \mathbf{k}_{v}^{(l)}]),
     \label{attention}
\end{align}
where $g$ is a trainable vector and $[\cdot \Vert \cdot]$ denotes the concatenation operation, $\tanh$ constrains the attention weight to $[-1,1]$, making \textsc{FaAt} adaptively serve as low- and high-pass filters. 

\textbf{Edge-Level Residual Connection.}
Inspired by \citet{realformer}'s findings that residual connections on attention coefficients can enhance performance, we incorporated residual connections into the attention coefficients of \ourmethod{} to expedite convergence. To elaborate, we introduced residual connections after obtaining the initial raw attention coefficients $\hat\alpha_{uv}^{(l)}$ by
\begin{align}
    \alpha_{uv}^{(l)} = (1-\beta)\hat\alpha_{uv}^{(l)} + \beta \alpha_{uv}^{(l-1)},
    \label{eq:beta1}
\end{align}
where $\beta \in [0,1]$ is a hyperparameter. 

\textbf{Heterogeneous Neighbor Aggregation.}
After getting the attention coefficients, we proceed to aggregate neighborhood messages with regard to the intrinsic heterogeneity in social networks. The aggregation paradigm of neighboring users for a central user, based on the relation type, is given as follows:
\begin{align}
    \mathbf{z}_{u}^{(l)} = \sum_{r \in \mathcal{R}} \sum_{v \in \mathcal{N}_r(u)} \frac{1}{ \sqrt{d_{ur}d_{vr}}}
    \alpha_{uv}^{(l)} \mathbf{W}_r^{(l)} \mathbf{x}_v^{(l-1)},
\end{align}
where $\mathbf{W}_r^{(l)}$ denotes learnable weight matrix for relation $r$, $d_{ur}$ and $d_{vr}$ respectively denote the degree of node $u$ and $v$ under relation type $r$, $\mathcal{R}$ denotes the set of relation type.

\textbf{Multi-head Attention.}
Inspired by the vanilla attention mechanism \cite{transformer}, we adopt multi-head attention to enhance the \ourmethod{}'s stability and performance. The corresponding update paradigm is:
\begin{align}
    \alpha_{uvk}^{(l)} = (1-\beta)\hat\alpha_{uvk}^{(l)} + \beta \alpha_{uvk}^{(l-1)},
    \label{eq:beta2}
\end{align}
\begin{align}
    \mathbf{z}_{u}^{(l)} = \Big\Vert_{k=1}^{K} \sum_{r \in \mathcal{R}} \sum_{v \in N_r(u)} \frac{1}{ \sqrt{d_{ur}d_{vr}}}
    \alpha_{uvk}^{(l)} \mathbf W_{rk}^{(l)} \mathbf{x}_v^{(l-1)},
\end{align}
where $K$ denotes the number of attention heads and $\Vert$ denotes concatenation operation.

\textbf{Node-Level Residual Connection.}
Previous research has shown that GNNs are often plagued by the notorious problems of over-smoothing and gradient vanishing \cite{li2018deeper, xu2018representation}. These issues can be mitigated through the well-designed preactivation node residual connections \cite{li2020deepergcn}. We propose two types of node residual connections in \ourmethod{}, which exhibit nuanced performance variations across different datasets. 
\begin{align}
    \mathbf{x}_u^{(l)} = \sigma (\mathbf W_{res}^{(l)}\mathbf{x}_u^{(l-1)} + \mathbf{z}_{u}^{(l)}),
    \label{res_connection1}
\end{align}
where $\mathbf{W}_{\text{res}}^{(l)}$ is a linear transformation to adapt user representations from the previous layer. 
\begin{align}
    \mathbf{x}_u^{(l)} = \sigma (\epsilon \mathbf{x}_u^{(0)} + (1 - \epsilon)\mathbf{z}_{u}^{(l)}),
    \label{res_connection2}
\end{align}
Equation \ref{res_connection2} adopts a weighted sum of initial user features to prevent over-smoothing, where $\epsilon \in [0,1]$ is a scaling factor to balance initial user features and learned representations.

\subsection{Training and Inference}
We apply $L$ layers of frequency adaptive attention module, and use a linear transformation to get the bot detection results:
\begin{align}
    \hat{y_u} = \mathbf{W_o} \mathbf{x}_u^{(L)} +\mathbf{b_o},
\end{align}
The attention coefficient $\alpha_{uv}$ plays a crucial role in the proposed \textsc{FaAt}. For edge $e_{uv} \in \mathcal{E}$ that connects nodes with the same label (homophilic), a positive attention coefficient (close to $1$) is desired to act as a low-pass filter. Conversely, for edges $e_{uv}$ which connects nodes with different labels (heterophilic), a negative attention weight (close to $-1$) is preferred to act as a high-pass filter. 
To ensure that the \textsc{FaAt} functions as intended, we incorporate an auxiliary loss derived from known labels:
\begin{align}
    \mathcal{L}_e = \frac{1}{\lvert \mathcal{E}_t \rvert} \sum_{(u,v) \in \mathcal{E}_t} \max(0, 1- \overline{\alpha_{uv}}\cdot y_{uv}),
\end{align}
where $\mathcal{E}_t$ represents the edge set that includes both user $u$ and $v$ in the training set, $\overline{\alpha_{uv}}$ denotes the average frequency adaptive attention coefficient across all \textsc{FaAt} layers. Furthermore, $y_{uv} = 1$ if $y_u=y_v$, otherwise, $y_{uv}=-1$.

To train \ourmethod{}, we integrate the cross entropy loss with the proposed weight guidance loss as the ultimate objective. Additionally, the $L_2$ regularization term is also added to alleviate overfitting. The resulting loss function can be expressed as:
\begin{align}
    \mathcal{L} = \sum_{u \in \mathcal{V}_t} y_u\log(\hat{y_u}) + \lambda_1 \sum_{\theta \in \Theta} \theta^2 +\lambda_2 \mathcal{L}_e
    \label{eq:lambda}
\end{align}
where $\mathcal{V}_t$ denotes the training user set, $\hat{y_{u}}$ is \ourmethod{}'s prediction, $\Theta$ encompasses all the trainable parameters of \ourmethod{}, and $\lambda_1$ and $\lambda_2$ are hyperparamters to balance three loss terms. 

\section{Experiments}

\subsection{Experiment Settings}

\begin{table*}[t]
  \centering
   \caption{Accuracy, binary F1-score, and balanced accuracy of Twitter bot detection systems on the three datasets. We run each method five times and report the average value as well as the standard deviation. Bold indicates the best performance, \underline{underline} the second best, and ‘-’ indicates that MGTAB-22 \cite{mgtab} doesn't provide desired features. * denotes that the results are significantly better than the second-best under the student t-test. \ourmethod{} significantly outperforms both the feature-, text- and graph-based baselines and competitive heterophilous baselines.}
  \begin{adjustbox}{max width=1\linewidth}
    \begin{tabular}{lccccccccc}
    \toprule[1.5pt]
    \multirow{2}[3]{*}{\textbf{Model}} & \multicolumn{3}{c}{\textbf{Cresci-15-H}} & \multicolumn{3}{c}{\textbf{TwiBot-20}} & \multicolumn{3}{c}{\textbf{MGTAB-22}} \\
    \cmidrule(r){2-4} \cmidrule(r){5-7} \cmidrule(r){8-10}      
    & Acc   & F1    & bAcc  & Acc   & F1    & bAcc  & Acc   & F1    & bAcc \\
    \midrule[0.75pt]
   \textsc{MLP}   & 88.19~\small{(1.0)} & 90.83 \small{(0.8)} & 86.66 \small{(1.1)} & 78.56~\small(1.1) & 81.71~\small(1.1) & 77.60~\small(0.9) & 86.46~\small(0.1) & 74.30~\small(1.0) & 82.09~\small(1.3) \\
   \textsc{SGBot} \cite{sgbot} & 96.30~\small(0.3) & 96.18~\small(0.2) & 95.70~\small(0.3) & 79.93~\small(0.5) & 83.89~\small(0.6) & 78.15~\small(0.3) & - & - & - \\
   \textsc{BotHunter} \cite{bothunter} & \underline{96.83}~\small(0.5) & \underline{97.48}~\small(0.4) & \underline{96.82}~\small(0.5) & 72.46~\small(0.8)	& 76.72~\small(0.9) & 71.30~\small(0.7)  & - & - & - \\
   \textsc{Roberta} \cite{roberta} & 96.41~\small(0.3) & 97.22~\small(0.2)& 95.84~\small(0.4) & 71.93~\small(0.8) & 	76.58~\small(0.4) & 70.59~\small(1.0)  & - & - & - \\
   \textsc{LOBO} \cite{lobo} & 96.01~\small(0.2) & 96.42~\small(0.2) & 95.98~\small(0.2) & 76.18~\small(0.2) & 	80.45~\small(0.3) & 74.66~\small(0.3)  & - & - & - \\
   \textsc{BotBuster} \cite{botbuster} & 96.68~\small(0.2) & 96.42~\small(0.2) & 96.28~\small(0.3) & 79.34~\small(0.8)	& 82.47~\small(1.2) & 78.30~\small(0.5)  & - & - & - \\
    \textsc{GCN} \cite{gcn}   & 75.21~\small(1.9) & 81.34~\small(2.2) & 71.36~\small(0.8) & 72.88~\small(0.4) & 76.41~\small(0.8) & 72.10~\small(0.3) & 85.78~\small(0.2) & 73.6~\small{ (0.8}) & 81.91~\small(0.9) \\
    \textsc{GraphSAGE} \cite{sage} & 81.44~\small(2.6) & 86.05~\small(1.4) & 78.29~\small(4.2) & 77.49~\small(0.3) & 80.45~\small(0.4) & 76.76~\small(0.3) & 85.22~\small(1.1) & 73.02~\small(1.2) & 81.69~\small(1.1) \\
    \textsc{GAT} \cite{gat}  & 93.92~\small(0.5) & 95.24~\small(0.4) & 93.10~\small(0.8) & 73.31~\small(0.5) & 76.81~\small(0.8) & 72.51~\small(0.5) & 86.79~\small(0.3) & 74.45~\small(1.0) & 81.91~\small(0.9) \\
    \textsc{BotRGCN} \cite{botrgcn} & 96.56~\small(0.3) & 97.07~\small(0.2) & 95.85~\small(0.5) & 77.85~\small(1.0) & 80.84~\small(1.3) & 77.06~\small(0.8) & 87.25~\small(0.4) & 75.42~\small(1.2) & 82.61~\small(1.1) \\
    \textsc{RGT} \cite{rgt}   & 96.63~\small(0.3) & 97.14~\small(0.2) & 96.07~\small(0.6) & 77.70~\small(0.9) & 80.68~\small(1.2) & 76.91~\small(0.7) & 87.47~\small(0.5) & 76.29~\small(2.1) & 83.56~\small(2.1) \\
    \textsc{BIC} \cite{bic} & 96.13
~\small(0.9) & 96.94
~\small(0.8) & 95.78~\small(1.2) & 75.75~\small(1.3) & 79.24~\small(1.8) & 74.77~\small(0.9) & - & - & - \\
    \hline
    \textsc{LINKX} \cite{linkx} & 91.65~\small(0.5) & 93.67~\small(0.4) & 89.51~\small(0.7) & 76.82~\small(1.8) & 81.35~\small(1.1) & 75.04~\small(2.4) & 87.42~\small(0.1) & 76.83~\small(1.1) & 84.25~\small(1.3) \\
    \textsc{MixHop} \cite{mixhop} & 89.80~\small(1.3) & 92.13~\small(1.1) & 88.12~\small(1.5) & 74.12~\small(0.3) & 77.18~\small(0.6) & 73.54~\small(0.4) & 86.92~\small(0.6) & 74.03~\small(1.7) & 81.33~\small(1.5) \\
    \textsc{H2GCN} \cite{h2gcn} & 95.15~\small(0.3) & 96.23~\small(0.2) & 94.22~\small(0.5) & 78.17~\small(0.7) & 81.17~\small(1.0) & 77.34~\small(0.4) & \underline{87.71}~\small(0.3) & \underline{77.49}~\small(0.6) & \underline{84.57}~\small(0.8) \\
    \textsc{FAGCN} \cite{fagcn} & 95.25~\small(0.8) & 96.31~\small(0.6) & 94.28~\small(1.1) & \underline{80.93}~\small(0.9) & \underline{84.10}~\small(1.1) & \underline{79.67}~\small(0.7) & 87.43~\small(0.3) & 77.02~\small(1.2) & 84.52~\small(1.5) \\
    \midrule[0.75pt]
    \ourmethod{} (Ours)  & \textbf{97.52*}~\small{(0.2)} & \textbf{98.06*}~\small{(0.1)} & \textbf{96.98*}~\small{(0.3)} & \textbf{82.29*}~\small{(0.2)} & \textbf{85.49*}~\small{(0.3)} & \textbf{80.82*}~\small{(0.2)} & \textbf{88.68*}~\small{(0.1)} & \textbf{79.21*}~\small{(0.6)} & \textbf{85.91*}~\small{(0.9)} \\
    \bottomrule[1.5pt]
    \end{tabular}%
    \end{adjustbox}
   
    \label{results}
\end{table*}%

\begin{figure*}[t]
    \centering
    \includegraphics[width=0.9\linewidth]{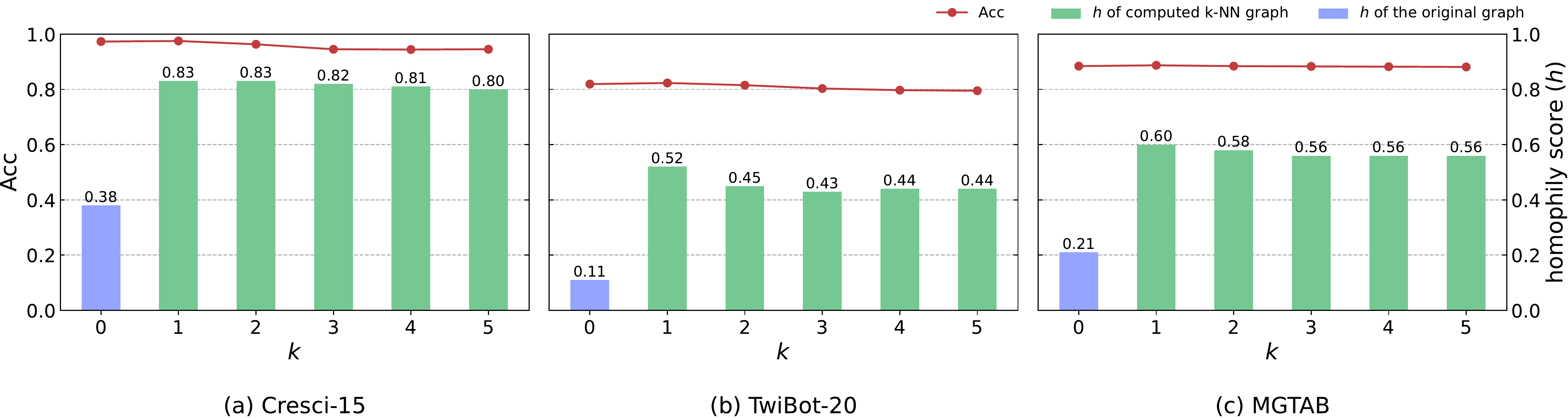}
    \caption{Augmentation Study: To investigate how parameter $k$ of \textsc{Homo-Aug} impacts the performance, we vary the value of $k$ and plot corresponding accuracy of \ourmethod{} and the homophily score of $k$-NN graph. }
    \label{dif_k}
\end{figure*}

\subsubsection{Datasets}
We evaluate \ourmethod{} on three widely-acknowledged Twitter bot detection benchmarks: Cresci-15 \cite{cresci15}, TwiBot-20 \cite{twibot20}, and MGTAB-22 \cite{mgtab}. 
The Cresci-15 dataset was proposed in 2015 and comprised 5,301 Twitter users and their follow relationships. We find it has a homophilic graph structure, as heterophilous disguise of novel bots was not conspicuous then. To replicate the heterophilous disguise, we curate the \textsc{Cresci-15-H} dataset by incorporating additional inter-class connections between humans and bots, while maintaining all other connections unaltered.
TwiBot-20 \cite{twibot20} dataset covers diversified bots and genuine users to better represent the real-world Twittersphere. MGTAB-22 \cite{mgtab} dataset contains 10,199 expert-annotated users and 7 types of relationships. Dataset statistics are presented in Table \ref{dataset stata}. Consistent with the approach taken by \citet{mgtab, shi2023over}, we have employed random splits with train/val/test ratios of $10\%/10\%/80\%$, and have reported the average performance over five runs, along with the standard deviation to ensure a fair comparison between \ourmethod{} and the baselines.

\subsubsection{Baselines}
We compare \ourmethod{} with Twitter bot detection baselines from feature-, text- and graph-based categories (MLP, SGBot, BotHunter, RoBERTa, LOBO, BotBuster, GCN, GraphSAGE, GAT, BotRGCN, RGT) as well as competitive heterophilous graph learning methods (LINKX, MixHop, H2GCN, FAGCN).  

\begin{itemize}[leftmargin=*]
\item \textbf{MLP}. A two-layer MLP that takes users' numerical, categorical, and description features as input.
\item \textbf{GCN} \cite{gcn} equally aggregates features from neighbors and pass user representations into an MLP for classification.
\item \textbf{GAT} \cite{gat} introduces the attention mechanism to distinguish the importance of neighboring users in aggregation. The learned representations are fed into an MLP for classification. 
\item \textbf{GraphSAGE} \cite{sage} separately embeds ego and neighbor user features and the learned representations are passed into an MLP for bot detection. 
\item \textbf{SGBot} \cite{sgbot} extracts features from users' metadata and exploits a random forest classifier for scalable and generalizable bot detection.  
\item \textbf{BotHunter} \cite{bothunter} extracts user features from metadata and text and presents a tiered approach to bot detection.
\item \textbf{RoBERTa} \cite{roberta} encodes user's descriptions and tweets using pretrained RoBERTa and feeds user features into an MLP for bot identification.
\item \textbf{LOBO} \cite{lobo} extracts features from user metadata and tweets and employs a random forest to identify diverse bots. 
\item \textbf{BotBuster} \cite{botbuster} enhances cross-platform bot detection by processing user metadata and textual information with the mixture-of-experts architecture. 
\item \textbf{BotRGCN} \cite{botrgcn} constructs a heterogeneous Twitter social network and exploits relational graph convolution networks for user representation learning and Twitter bot detection. 
\item \textbf{RGT} \cite{rgt} models the inherent heterogeneity in the Twittersphere with relational graph transformers to enhance Twitter bot detection. 
\item \textbf{BIC} \cite{bic} proposes a text-graph interaction module and models semantic consistency thus improving bot detection performance and combating bot evolution.
\item \textbf{LINKX} \cite{linkx} is a simple and scalable method that separately embeds node features and adjacent matrix and feeds them into MLPs for heterophilous graph representation learning. 
\item \textbf{MixHop} \cite{mixhop} utilizes multi-hop neighbors to get more homophilic information, thereby enhancing heterophilic graph learning. 
\item \textbf{H2GCN} \cite{h2gcn} separately embeds ego and higher-order neighborhood features and shows strong performance on heterophilous benchmarks. 
\item \textbf{FAGCN} \cite{fagcn} employs frequency adaptive filters through the computation of edge-level aggregation weights that may have negative values.

\end{itemize}

\begin{table}[t]
    \centering
   \caption{Statistics of the three Twitter bot detection datasets: Cresci-15 \cite{cresci15}, TwiBot-20 \cite{twibot20}, and MGTAB-22 \cite{mgtab}, featuring increasingly up-to-date collections of Twitter as well as decreasing homophily score, verifying the heterophilous disguise phenomenon.}
    \begin{adjustbox}{max width=1\linewidth}
    \begin{tabular}{l|cccc}
    \toprule[1.5pt]
    \textbf{Dataset} & \textbf{\# user} & \textbf{\# human} & \textbf{\# bot} & \textbf{$homophily$-score} \\
    \midrule[0.75pt]
    \textsc{Cresci-15} \cite{cresci15} & 5,301 & 1,950 & 3,351 & 0.98\\
    $\hookrightarrow$ \textsc{Cresci-15-H} & 5,301 & 1,950 & 3,351 & 0.38\\
    \textsc{TwiBot-20} \cite{twibot20} & 229,580 & 5,237 & 6,589 & 0.11\\
    \textsc{MGTAB-22} \cite{mgtab} & 410,199 & 7,451 & 2,748 & 0.21 \\
    \bottomrule[1.5pt]
    \end{tabular}
    \end{adjustbox}
 
    \label{dataset stata}
\end{table}

\subsubsection{Implementation}
We use PyTorch \cite{pytorch}, PyTorch Geometric \cite{pyg}, Scikit-learn \cite{scikit}, and Numpy \cite{numpy} to implement \ourmethod{}. The hyperparameter settings are presented in Table \ref{hyperpram} to facilitate reproduction. We conduct all experiments on a cluster with 4 RTX3090 GPUs with 24 GB memory, 10 CPU cores, and 64 GB CPU memory. Following \citet{rgcn, rgt}, we leverage follower and friend relationships to construct the heterogeneous information network. For user features, we use numerical properties, categorical properties, and descriptions in Cresci-15 \cite{cresci15} and TwiBot-20 \cite{twibot20}. Unlabeled support set is not used in TwiBot-20 and MGTAB-22. It takes about 0.5 minute, 1 minute, and 2 minutes to train \ourmethod{} on Cresci-15, TwiBot-20, and MGTAB-22. We will make all \ourmethod{} implementations publicly available.

\begin{table}[t]
    \centering
    \caption{Hyperparameter settings of \ourmethod{}}
    \begin{adjustbox}{max width=1\linewidth}
    \begin{tabular}{l|ccc}
    \toprule[1.5pt]
    \textbf{Settings} & \textbf{Cresci-15-H} & \textbf{TwiBot-20} & \textbf{MGTAB-22}\\
    \midrule[0.75pt]
    \textsc{Optimizer} & Adam & Adam & Adam \\
    \textsc{learning rate} & $0.001$ & $0.0005$ & $0.001$ \\
    \textsc{L2 regularization} $\lambda_1$ & $0.01$  & $0.01$  & $0.001$ \\
    $\lambda_2$ (\textsc{Eq} \ref{eq:lambda})& 0.1   & 0.2   & 0.1 \\
    $\beta$ (\textsc{Eq} \ref{eq:beta1}, \ref{eq:beta2}) & 0.1   & 0.1   & 0.1 \\
    \textsc{dropout} & 0.3   & 0.4   & 0.1 \\
    \textsc{num heads} $K$& 4     & 4     & 4 \\
    \textsc{hidden size} & 128   & 128   & 64 \\
    \textsc{layer} $L$ & 2     & 2     & 5 \\
    $k$ (\textsc{Eq} \ref{eq:knn})   & 1     & 1     & 1 \\
    \textsc{Residual Connection} & \textsc{Eq} \ref{res_connection1} & \textsc{Eq} \ref{res_connection2} & \textsc{Eq} \ref{res_connection2}\\
    \bottomrule[1.5pt]
    \end{tabular}
    \end{adjustbox}

    \label{hyperpram}
\end{table}

\begin{figure*}[t]
    \centering
    \includegraphics[width=0.95\linewidth]{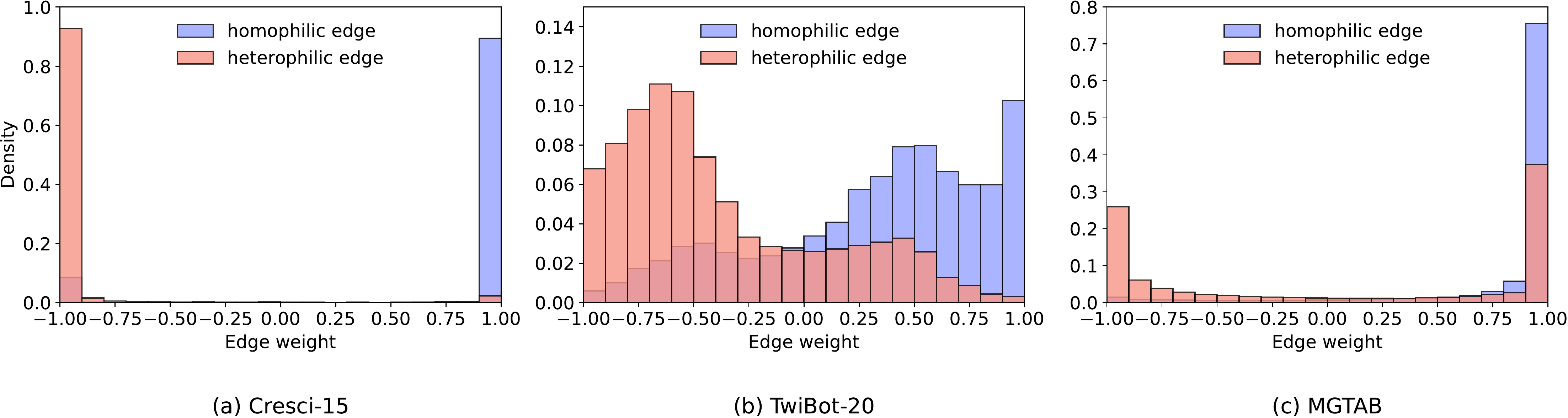}
    \caption{Visualization of mean attention coefficients extracted from the first layer of \textsc{FaAt} on different datasets. }
    \label{fig:weight_distribution}
\end{figure*}

\subsection{Main Results}
We evaluate \ourmethod{} with 16 representative feature-, text-, graph-based Twitter bot detection methods, as well as competitive heterophilous GNNs. As presented in Table \ref{results}, we find:

\begin{itemize}[leftmargin=*]
\item \ourmethod{} consistently outperforms all baseline methods across all three datasets. Specifically, compared with previous state-of-the-art Twitter bot detection method RGT \cite{rgt}, \ourmethod{} achieves 4.59\%, 4.81\%, 3.91\% improvements on accuracy, F1-score, and balanced accuracy respectively on TwiBot-20 \cite{twibot20}. Similarly, on MGTAB-22 \cite{mgtab}, \ourmethod{} outperform RGT by 1.21\%, 2.92\%, and 2.35\% on accuracy, F1-score, and balanced accuracy, respectively. Compared with state-of-the-art heterophilic graph learning method FAGCN, \ourmethod{} achieves 1.36\% accuracy and 1.39\%  F1-score higher on TwiBot-20, and 1.25\% accuracy and 2.19\% F1-score higher on MGTAB-22.

\item Existing homophilic graph-based Twitter bot detection methods generally show inferior performance compared with heterophilous graph learning methods. Specifically, previous state-of-the-art graph-based method RGT \cite{rgt} has been found to be inferior to vanilla feature- and text-based methods on the Cresci-15-H dataset. Furthermore, our experiments demonstrate that a simple MLP can outperform GCN on all three datasets, highlighting the significant impact of heterophilous disguise on graph-based Twitter bot detection methods.
\item As graph heterophily increases, the superiority of \ourmethod{} over baselines increases as well. 
Specifically, \ourmethod{} achieves highest improvements (4.81\% and 3.91\%) on F1-score and balanced accuracy higher than RGT on TwiBot-20 with the lowest homophily, demonstrating a higher efficiency in detecting bots and the effectiveness of \ourmethod{} in solving heterophilous disguise challenge. 
\end{itemize}

\begin{figure}[t]
    \centering
    \includegraphics[width=1\linewidth]{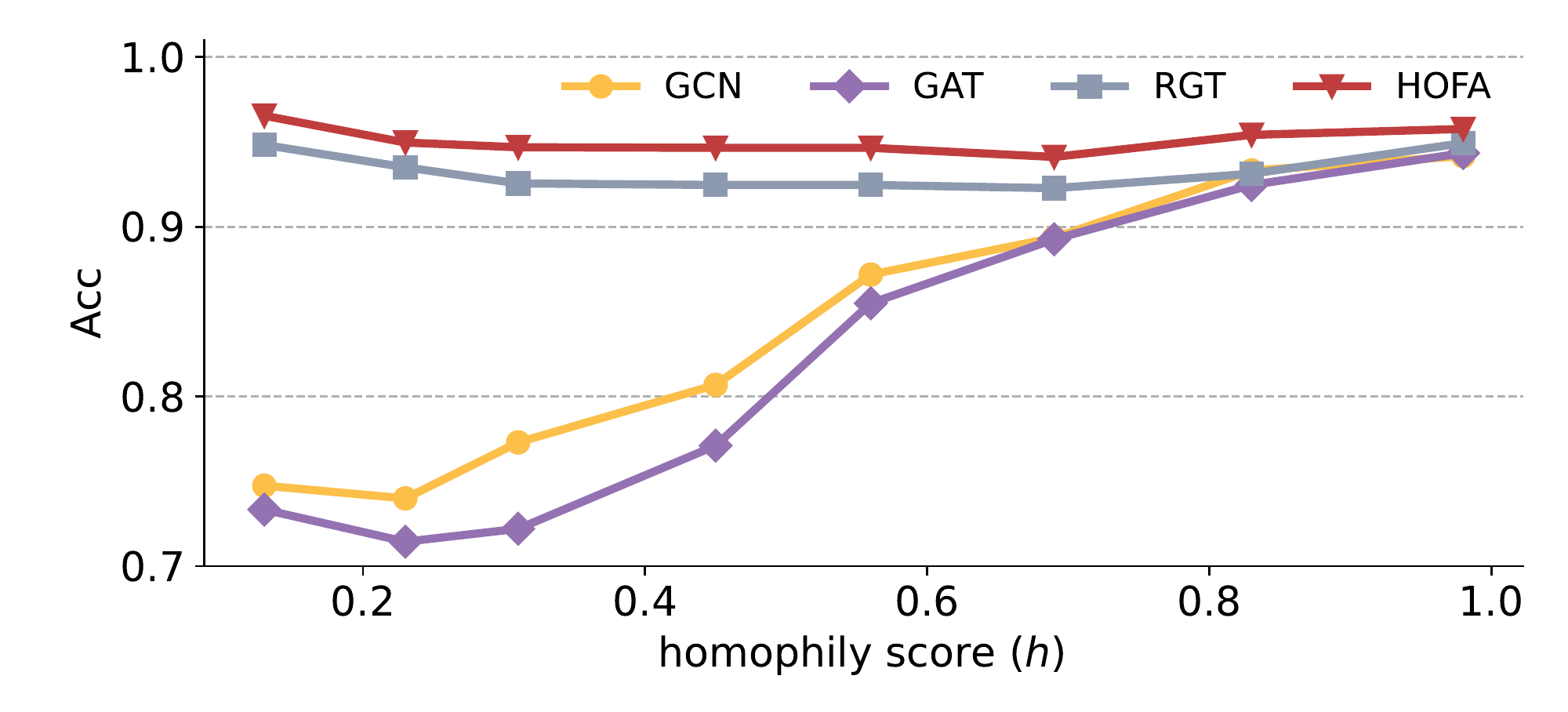}
    \caption{Heterophily Perturbation Study: To investigate how heterophily impacts bot detection performance, we change the homophily score of Cresci-15 by introducing heterophilic edges and we test GCN, GAT, RGT, and \ourmethod{}'s performance under different homophily settings. }
    \label{fig:change_hetero}
\end{figure}

\begin{figure*}[t]
    \centering
    \includegraphics[width=0.9\linewidth]{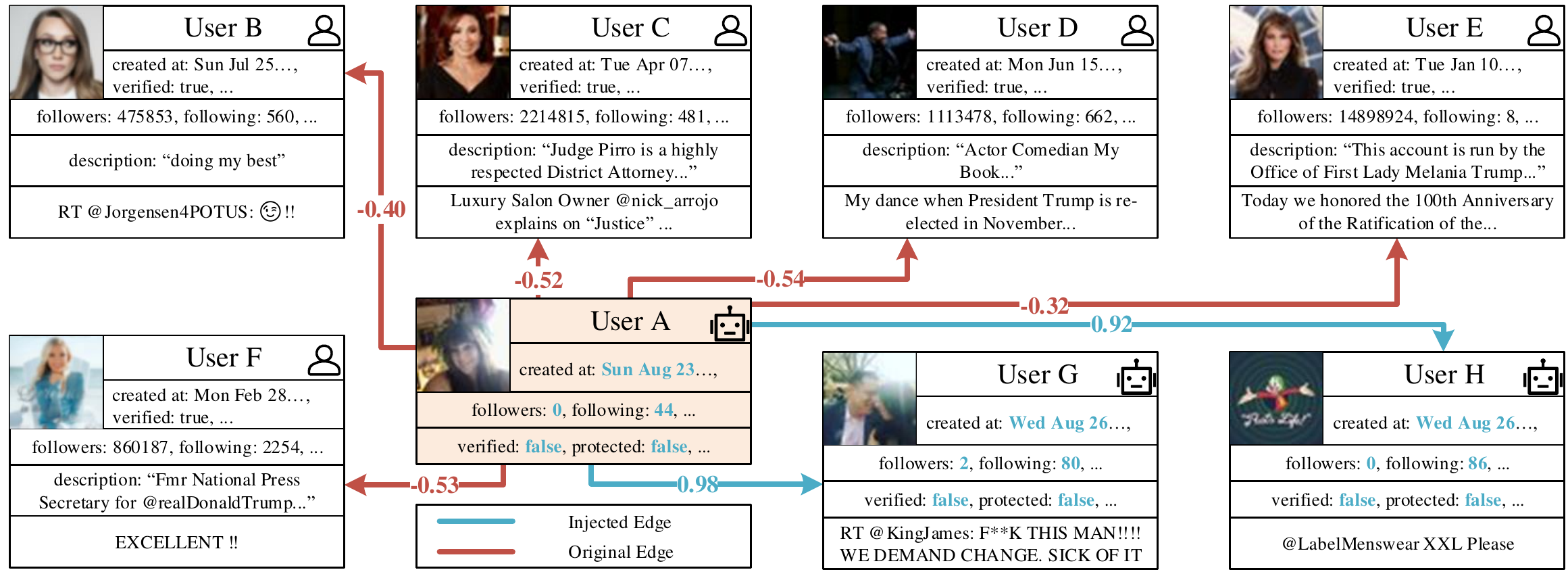}
    \caption{Case Study: we display a typical heterophilous disguised bot following many genuine users. We show another two bot accounts discovered by \textsc{Homo-Aug} module and visualize the attention weights between them.}
    \label{fig:case_study}
\end{figure*}

\subsection{Augmentation Study}
To investigate the effectiveness of our homophily-oriented graph augmentation module and the impact of the parameter $k$ on our model's performance, we vary the value of $k$ in the \textsc{Homo-Aug} module and examine the homophily score of the injected $k$-nearest neighbor ($k$-NN) graphs and the corresponding bot detection accuracy of \ourmethod{}. As presented in Figure \ref{dif_k}, we find that the homophily score of computed $k$-NN graph is significantly larger than the original graph, indicating \textsc{Homo-Aug}'s effectiveness in improving the homophilic information and thereby enhancing model's performance. However, as the value of $k$ increased, both the homophily of the $k$-NN graph and the accuracy of \ourmethod{} decreased. We speculate that this is due to the introduction of more noise into the graph structure with higher values of $k$.

\subsection{Heterophily Pertubation Study}
As indicated by the main results, the superiority of \ourmethod{} increases with decreasing homophily scores, to further investigate the impact of heterophily on Twitter bot detection performance, we modify the graph structure of the Cresci-15 \cite{cresci15} dataset by introducing different proportions of heterophilic edges while maintaining a constant number of homophilic edges. This resulted in graphs with varying homophily scores. We evaluated the performance of GCN \cite{gcn}, GAT \cite{gat}, RGT \cite{rgt}, and our proposed \ourmethod{} on these graphs with different homophily scores, and demonstrate results in Figure \ref{fig:change_hetero}. Experimental results demonstrate that heterophily can have a significant impact on the performance of homophilic GNNs, e.g., the performances of both GCN and GAT degrade as the homophily score decreases. In contrast, the previous state-of-the-art method, RGT \cite{rgt}, which is based on the relational transformer architecture, exhibits relatively stable performance as the homophily varies. Our proposed method, \ourmethod{}, consistently achieves the best performance regardless of whether the graph is homophilic or heterophilic, thus verifying \ourmethod{}'s robustness against different extents of heterophily and its effectiveness in combating the heterophilous disguise challenge.

\subsection{Visualization of Attention Coefficients} 
To examine our proposed \ourmethod{} can learn different attention coefficients along different types of edges, as intended, we visualize the distribution of mean attention coefficients $\alpha_{uv}$ extracted from the first layer in Figure \ref{fig:weight_distribution}. To improve readability, we categorize the edges into homophilic edges and heterophilic edges based on whether the two connected nodes have the same ground truth label. Our results demonstrate that the majority of heterophilic edges have negative attention coefficients, while the majority of homophilic edges have positive attention coefficients, which aligns with our design intention of using a low-pass filter (with positive attention weight) along a homophilic edge and a high-pass filter (with negative attention weight) along a heterophilic edge.

\subsection{Case Study}
To enhance the understanding of \ourmethod{}, we conducted a case study on a specific disguised bot account by analyzing its adaptive attention weights and the injected $k$-NN edges. 
As shown in Figure \ref{fig:case_study}, the heterophilous disguised bot mostly follows genuine users. Our \textsc{Homo-Aug} module identifies another two bot accounts that closely resembles the disguised bot. The attention weights, computed by our proposed \textsc{FaAt} module, reveal that the learned weights between a human and a bot are negative, indicating that our module automatically selects a high-pass filter to accentuate the dissimilarities in representation between humans and bots. Conversely, the attention weights between two bot accounts are positive, indicating that a low-pass filter is employed to smooth representations for these two users. These findings, together with our results, corroborate the effectiveness of \ourmethod{} to combat the heterophilous disguise challenge in Twitter bot detection.


\subsection{Ablation Study}
As \ourmethod{} significantly outperforms all baseline methods across the three benchmark datasets, we conduct further analysis on TwiBot-20 and MGTAB-22 dataset to examine the impact of our design choice. The results are present in Table \ref{ablation}. 
We examine the role of \textsc{Homo-Aug}, \textsc{FaAt}, and weight guidance loss $\mathcal{L}_e$ in \ourmethod{}. The accuracy decreases by 0.36\% on TwiBot-20 and the accuracy and F1-score decreases by 0.25\% and 1.13\% on MGTAB-22 after removing the \textsc{Homo-Aug} module. We replace the \textsc{Homo-Aug} module with other augmentation method, \textit{i.e.}, adding random edges, and the performance further degrades. We also replace the \textsc{FaAt} with vanilla attention mechanism and mean pooling aggregation. The results demonstrate a significant decrease in model's accuracy and F1-score, highlighting the essential role of proposed \textsc{FaAt}. We remove the weight guidance loss (\textsc{WeGL}) in training stage. We find that accuracy and F1 score decreases 0.96\%, 0.95\% on TwiBot-20 and 0.3\%, 0.98\% on MGTAB-22, illustrating the role of \textsc{WeGL} in guiding the \textsc{FaAt} module. 
Together with these ablated and full model performances, we prove the effectiveness of our design choice.

\begin{table}[t]
    \centering
  \caption{Ablation Study of \ourmethod{}. MP denotes mean pooling.}
    \begin{tabular}{cccccc}
    \toprule[1.5pt]
    \multirow{2}[3]{*}{\textbf{Settings}} &  \multicolumn{2}{c}{\textbf{TwiBot-20}} & \multicolumn{2}{c}{\textbf{MGTAB-22}} \\
    \cmidrule(r){2-3} \cmidrule(r){4-5} & Acc & F1  & Acc & F1  \\
     \midrule[0.75pt]
     full model & \textbf{82.29} & \textbf{85.49} & \textbf{88.68} & \textbf{79.21} \\
     \midrule
     w/o augmentation & \underline{81.93} & \underline{85.23} & \underline{88.43} & 78.08 \\
     add random edge & 81.90 &  85.17 & 88.32 & 78.09\\
     w/o frequency adaptive & 81.30 & 84.53 & 87.18 & 75.05\\
     replace \textsc{FaAt} with MP & 78.02 &   80.91 & 87.34 & 75.66 \\
     w/o \textsc{WeGL} & 81.33 & 84.54 & 88.38 & \underline{78.23} \\
    \bottomrule[1.5pt]
    \end{tabular}
    \label{ablation}
\end{table}

\section{Discussion}
\subsection{Ethical Statement}
We envision \ourmethod{} as an aid rather than the sole ultimate decision-maker in detecting Twitter bots. Human oversight is necessary for making final judgments as \ourmethod{} and other automatic bot detection frameworks are imperfect proxies and thus need to be used with care. First, \ourmethod{} can have false positive predictions, \emph{i.e.}, misclassifying legitimate accounts into bots, which can bring troubles to businesses or individuals who spread important information \cite{thieltges2016devil}. Second, \ourmethod{} leverage pretrained language model (PLM) for user feature encoding and GNN for user representation learning, which would inevitably inherit bias and stereotype from PLM and GNN. 
For example, a significant portion of pertaining data for PLMs contains hate, bias, and stereotype \cite{lmbias, nadeem2020stereoset, feng2023pretraining}. In addition, GNNs may inherit historical prejudices from training data and lead to discriminatory bias in predictions \cite{wang2022improving, dong2022edits}.

\subsection{Limitations and Future Work}
We identify two limitations in \ourmethod{}. First, it operates on a two-stage training method, where an MLP is initially trained for graph data augmentation, then the augmented graph and node are fed into \textsc{FaAt} for bot identification. This two-stage training paradigm may result in sub-optimal performance. Second, due to limited data sources, we are only able to detect bot accounts on Twitter, while \ourmethod{}'s generalization to other social platforms, such as Reddit, Facebook, and TikTok remains under-explored. Moving forward, we plan to develop a universal bot detection framework with self-supervised learning that can adapt to diverse social platforms.

\section{Conclusion}
Detecting Twitter bots is a crucial yet arduous task. However, we find that Twitter bots tend to follow genuine users to disguise themselves, namely heterophilous disguise challenge. To tackle this challenge, we propose a novel Twitter bot detection framework \ourmethod{}, that comprises a homophily-oriented graph augmentation module (\textsc{Homo-Aug}) and a frequency adaptive attention module (\textsc{FaAt}). To be more specific, we first augment the graph data with \textsc{Homo-Aug} to improve the graph homophily, we then feed user features into \textsc{FaAt} layers to adaptively learn low- and high-pass filters for homophilic and heterophilic edges, respectively. Our experiments demonstrate that \ourmethod{} achieves state-of-the-art performance on three Twitter bot detection benchmarks. Additional studies further illustrate the effectiveness of our proposed \textsc{Homo-Aug} and \textsc{FaAt} module, and highlight \ourmethod{}'s ability to address the heterophilous disguise challenge in Twitter bot detection. 

\bibliographystyle{ACM-Reference-Format}
\bibliography{main}

\appendix

\end{document}